\documentclass{aa}  
\usepackage{graphicx}
\usepackage{txfonts}
\usepackage{natbib}
\newcommand{\sgr}{\object{Sgr dSph}}
\newcommand{\mfifo}{\object{M 54}}
\newcommand{\ters}{\object{Terzan 7}}
\newcommand{\tere}{\object{Terzan 8}}
\newcommand{\arpt}{\object{Arp 2}}

\begin{document}
\title{A wide angle view of the Sagittarius dwarf Spheroidal Galaxy. \\ 
I: VIMOS photometry and radial velocities across \sgr~ major and minor axis.
\thanks{Based on observations made with ESO Telescopes at 
Paranal Observatory under programme ID 073.B-0455} }

\subtitle{}

\author{G. Giuffrida\inst{1} \and
        L. Sbordone\inst{2,3,4} \and
        S. Zaggia\inst{5} \and         
        G. Marconi\inst{6}     \and
        P. Bonifacio\inst{2,3,7} \and
        C. Izzo\inst{8} \and
        T. Szeifert\inst{6}    \and
        R. Buonanno\inst{9} 
         }
   \offprints{G. Giuffrida}

\institute{
	ASI Science data Center, Via Galileo Galilei, 00044, Frascati, Italy \\
	\email{giuffrida@mporzio.astro.it}
	\and
	CIFIST Marie Curie Excellence Team
	\and
	GEPI, Observatoire de Paris, CNRS, Universit\'e Paris Diderot; Place
          Jules Janssen 92190
          Meudon, France
    \and
    Max-Planck-Institut f\"ur Astrophysik, Karl-Schwarzschild-Str. 1, 
    Postfach 1317, 85741 Garching b. M\"unchen, Germany    
	\and
	INAF -- Osservatorio Astronomico di Padova
	Vicolo dell'Osservatorio 5, - 35122 - PADOVA - ITALY, Italy
	\and
	European Southern Observatory, 
	Alonso de Cordova 3107
          Vitacura
          Casilla 19001
          Santiago, Chile
	\and
	Istituto Nazionale di Astrofisica,
         Osservatorio Astronomico di Trieste,  Via Tiepolo 11,
          I-34143 Trieste, Italy
	\and
	European Southern Observatory, 
	Karl-Schwarzschild-Str. 2 
         D-85748 Garching bei M\"unchen
	\and
	Universit\'a di Roma Tor Vergata, 
	Via della Ricerca Scientifica, 1
         00133 Rome, Italy
	}

   \date{Received September 15, 1996; accepted March 16, 1997}

\abstract 
{The \object{Sagittarius dwarf Spheroidal Galaxy} (\sgr) provides us
  with a unique possibility of studying a dwarf galaxy merging event
  while still in progress.  Moving along a short-period, quasi-polar
  orbit in the Milky Way Halo, \sgr~ is being tidally dispersed
  along a huge stellar stream.  Due to its low distance (25 kpc), the
  main body of \sgr~ covers a vast area in the sky (roughly $15 \times
  7$ degrees). Available photometric and spectroscopic studies have
  concentrated either on the central part of the galaxy or on the
  stellar stream, but the overwhelming majority of the galaxy body has
  never been probed. }
{The aim of the present study is twofold. On the one hand, to produce
  color magnitude diagrams across the extension of \sgr~ to study its
  stellar populations, searching for age and/or composition gradients
  (or lack thereof). On the other hand, to derive spectroscopic
  low-resolution radial velocities for a subsample of stars to
  determine membership to \sgr~ for the purpose of high resolution
  spectroscopic follow-up.  }
{We used VIMOS@VLT to produce V and I photometry on 7 fields across
  the \sgr~ minor and major axis, plus 3 more centered on the
  associated globular clusters \ters, \tere~ and \arpt. A last field
  has been centered on \mfifo, lying in the center of \sgr. VIMOS high
  resolution spectroscopic mode has then been used to derive radial
  velocities for a subsample of the observed stars, concentrating on
  objects having colors and magnitudes compatible with the \sgr~ red giant
  branch.}
{We present photometry for 320,000 stars across the main body of \sgr,
  one of the richest, and safely the most wide-angle sampling ever
  produced for this fundamental object. We also provide robust
  memberships for more than one hundred stars, whose high resolution
  spectroscopic analysis will be the object of forthcoming papers. 
  \sgr~appears remarkably uniform among the observed fields. We confirm
the presence of a main \sgr~ population characterized roughly by the
same metallicity of 47 Tuc, but we also found the presence of multiple populations 
on the peripheral fields of the galaxy, with a metallicity spanning from [Fe/H]=-2.3 to a nearly
solar value.  }
{}
	
\titlerunning{VIMOS photometry of \sgr}


\maketitle
%

\section{Introduction}

Dwarf galaxy mergers are believed to play a key role in the buildup of
the Milky Way (MW) Halo \citep[][]{searle78,carollo07,diemand07}. These
are also invoked as suspect responsible of the appearance of the Thick
Disk \citep[][]{kroupa02}, of the Bulge/Bar system
\citep[][]{heller07}, and of the Disk warp \citep[][]{ibata97}. The
discovery of the \object{Sagittarius dwarf Spheroidal galaxy}
\citep[\sgr;][]{ibata94,ibata95} allowed for the first time to observe
the MW in the act of tidally accreting a dwarf galaxy. Immediately after
the discovery, theoretical predictions \citep[][]{ibata97,helmi99} as
well as observational hints
\citep[][]{ibata97,Ng1,ngs97,Ng2,majewski99,ivezic00} concurred in
indicating that \sgr~ should be releasing its stellar content in the
Halo along a massive, kinematically cold stream. The existence of the
stream was finally confirmed by  \citet[][]{ibata01} and \citet[][]{majewski03},
and further branches of it were also identified in the stellar sample
of the Sloan Digital Sky Survey
\citep[][]{newberg03,martinez04,belokurov06}. As a matter of fact, the \sgr~
stream is clearly the most prominent feature visible in the galactic
Halo.

All such discoveries increased the importance of \sgr~ not only as a
``test case'' of tidal merging, but as the source of a significant
portion of the stellar population in the Halo. \citet{majewski03}
estimated that roughly 75\% of high-latitude Halo M-giants originated
from \sgr~ debris, which also provided almost all Halo AGB C-stars
\citep[see][]{mauron05}. \citet{zijlstra06} estimated that up to 10\%
of the Halo material could come from \sgr~ debris. A thorough
reconstruction of the \sgr~ stream(s), moreover, would pose very
strong constraints on the  shape of the MW dark matter Halo
\citep[e.g.][]{fellhauer06}.
With the discovery of \sgr, 4 globular clusters (GC) which were before
considered belonging to the halo have been recognized as being
associated with the \sgr. 
While three of them (\ters, \tere, and \arpt) lay at the outskirts of \sgr, 
the position of the fourth one, \mfifo,  coincides with the center of the dwarf spheroidal \citep[][]{monaco05b}
although it likely formed elsewhere and fell subsequently into the galaxy
core \citep[][]{bellazzini08}. 
It became subsequently clear that a number of other globular clusters, currently
lying far from \sgr, have kinematical properties compatible with an
origin within the \sgr~ system, and a subsequent stripping
\citep[][]{bellazzini03}, or were actually embedded into the stream
material \citep[][]{martinez02}. For at least one of them, \object{Pal
  12}, the origin within the \sgr~ system has now been firmly
established \citep[][]{cohen04,sbordone2007}.
The large \sgr~size (roughly 15$\times$7 square degrees in the sky), a
consequence of its modest distance \citep[25 kpc, see][]{monaco04}, is
likely the reason why only the central region of the galaxy, and the
associated globular clusters, have been studied so far. \mfifo~ and the
surrounding area have been examined photometrically
\citep[e.g.][]{M98,bella99,bellazzini99B,LS00,monaco04,bella06, siegel07} as
well as by means of high resolution spectroscopy \citep[]{bonifacio00,bonifacio2004,tarantella04,zaggia2004,sbordone2005,caffau05,monaco2005,mcwilliam05,sbordone2007,mottini08}.
While ground based photometry suggests for the \sgr~
main body population a mean metallicity of [Fe/H]$\sim$-1, spectroscopy
shows a surprising abundance spread in the main body, with a
significant population around [Fe/H]=0, a mean value around
[Fe/H]=-0.5 and a weak metal poor tail possibly extending below
[Fe/H]=-2.5. 
The spectroscopic discovery of the metal rich population in the \sgr~core was then confirmed 
by the detection of its Turn Off superimposed over the M54 CMD in the HST ACS photometry by \citep[][]{siegel07}.
Globular clusters considered belonging to \sgr ~ spans
a similar wide range in metallicity, from [Fe/H]=-0.6 for \ters~ to [Fe/H]=-2.3 for
\tere. 

Stars in the \sgr~ stream appear to be slightly less metal rich
than those in the main body, with metallicity gradients  along the stream 
\citep[][]{bellazzini06b,clewley06,monaco07,chou07}.
This would not be unexpected if the most metal poor stars were located in the 
outskirts of \sgr~ and then subject to the tidal interaction with the MW.

\begin{table}[tp]\tiny
\caption{List of observed fields}
\label{radec}      
\centering          
\begin{tabular}{lcccccc} 
\hline\hline       
&&&&&&\\[-5pt]
Field & Pos. & RA        & DEC       & l      & b       & VIMOS \\ 
Name  &      & (J2000.0) & (J2000.0) &  (deg) & (deg)   & cameras \\[5pt]
\hline
&&&&&&\\[-5pt]
  M 54 & Cen & 18:55:03.3 & -30:28:41 &  5.954 &  -14.692 & 4 \\
  Sgr0 & Min & 18:56:12.9 & -30:01:48 &  6.482 &  -14.743 & 4 \\
  Sgr1 & Maj & 18:43:00.0 & -28:58:34 &  6.258 &  -11.689 & 4 \\
  Sgr2 & Maj & 18:48:56.1 & -29:41:49 &  6.139 &  -13.165 & 3 \\
  Sgr3 & Maj & 19:00:57.6 & -31:12:00 &  5.779 &  -16.140 & 3 \\
  Sgr4 & Maj & 19:07:00.0 & -31:57:04 &  5.561 &  -17.615 & 3 \\
  Sgr5 & Min & 18:57:13.6 & -29:35:24 &  6.990 &  -14.770 & 3 \\
  Sgr6 & Min & 18:52:44.6 & -31:20:31 &  4.931 &  -14.581 & 4 \\[5pt]
\hline
\end{tabular}
\end{table}

The detailed chemical analysis of \sgr~ population showed a
highly peculiar abundance pattern, with anomalous abundances for a
number of elements 
\citep[][and references therein]{sbordone2007}. The detection of the
same signature in \object{Pal 12} was the concluding argument to
 assign the cluster to the \sgr~ system.

Although it seems now established
\citep[e. g.][]{lanfra07,lanfra06A,lanfra06B} that galactic winds triggered by starbursts are
required to explain the low [$\alpha$/Fe] ratios
observed in many dSph, the complete history of the  \sgr~ is far from being clarified.
For instance, the extent to which \sgr~ has been altered by the
interaction with the MW is currently difficult to assess. The original mass of \sgr~ is
uncertain, as well as its original gas content. The large mass of the
stream, the unusually high metallicity among the dSph, and the large
population of GCs, are all indications that we are dealing with an object which was in the past more a dwarf elliptical than a ``classic'' dwarf
spheroidal galaxy.

There is thus a number of open questions regarding this 
complex stellar system, some of which can only be addressed by the analysis of 
stellar populations {\em across} the
main body of \sgr.
Among these open questions we will only  mention the following: how did star formation proceed in \sgr? 
How uniform is its chemical composition? 
Which is the lower limit of metallicity in \sgr and 
does it show gradients 
in composition or age ? What is the dynamics of the main body?

\begin{figure*}[t]
\centering
\includegraphics[width=18.5cm]{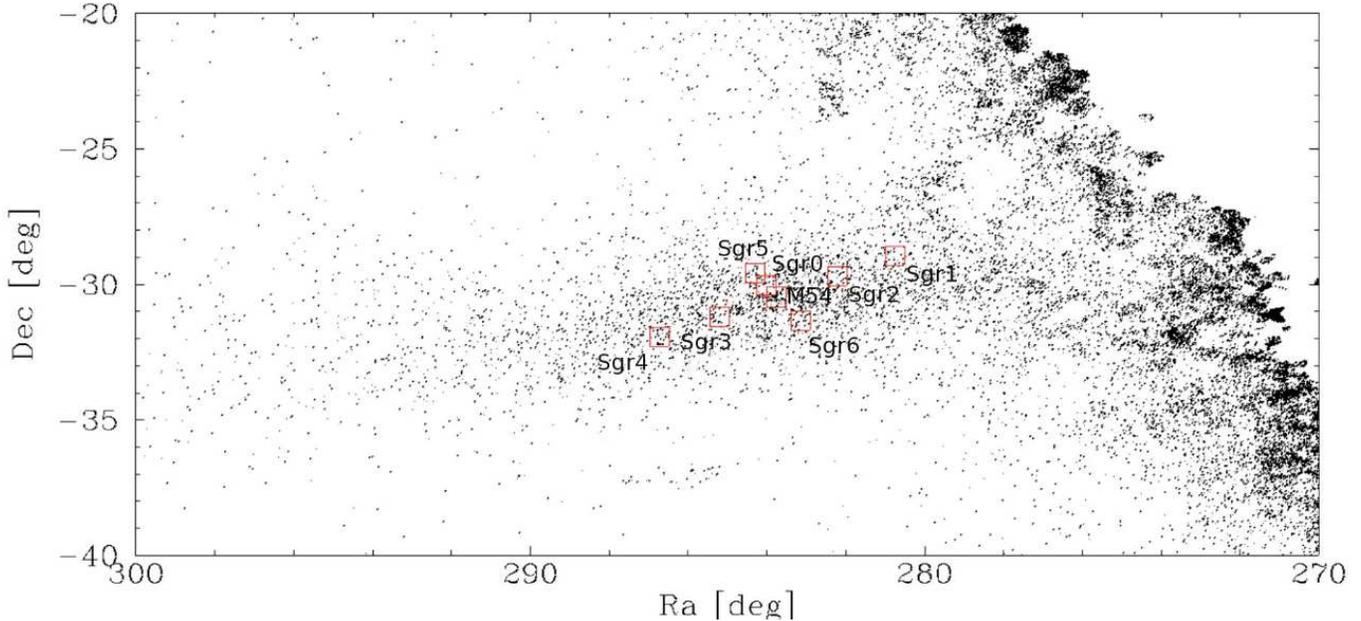}
   \caption{ Map of the \sgr~ galaxy obtained from 2MASS and UCAC catalog selected in K, J$-$K, 
   and E(B$-$V) (E(B$-$V) $< 0.555$, $0.95 <$ (J$-$K)$_0 < 1.10$, $10.5 <$ K$_0< 12$, see \citet{majewski03}. 
   The boxes shows the position of our fields}
      \label{mappa}
\end{figure*}


\section{Observations}

The program to sample the stellar populations across \sgr~ was 
performed in three steps. 

First, we secured
VIMOS@VLT images of 7 fields across the major and minor axis of the galaxy,
and of 5 fields for the associated GCs. 
Second,
we selected targets to observe with the VIMOS Multi Object
Spectroscopy (MOS) in the High Resolution red grism (645-860nm,
R=2500) mode, with an exposure time of 600s for each pointing.  Finally, we performed a follow-up
spectroscopical analysis with FLAMES \citep[GIRAFFE/UVES mode
see][]{pasquini00} on candidate \sgr~members in the 7 fields. 

This paper presents the photometric
study and the complete source catalog (available online) for the \sgr~ main body 
fields. The results for \ters,
\tere, \arpt~ and \object{NGC 4147} will be presented in a forthcoming
paper \citep[][]{giuffrida09}, although the fiducial lines derived
from that photometry will be employed here.  The abundance
and dynamical analysis from the high resolution spectra will be
the object of a further paper.

Observations with VIMOS@VLT imaging camera
\citep[][]{lefevre03} were obtained during 5 nights on March, April and May 2004.
The instrument consists of four channels, each with a field of view of 7 x 8 arcmin 
and a pixel size of 0.205 arcsec, with a gap of about 2 arcmin between each quadrant.

Each pointing consists of two 10s exposures in the V and in the I band; the
FWHM was in the range 0.7 to 1.3 arcsec.  Tab.~\ref{radec} presents
the list of the observed fields. Last column indicates the number of channels used for each field
because one of the VIMOS channels, Q1, turned out to be out of
focus on the night 18-04-2004. Consequently for all fields observed that night, Sgr2, Sgr3, Sgr4 and Sgr5,
only three channels, namely Q2, Q3 and Q4, could be used for photometry.
The position of the fields is reported in Fig. \ref{mappa},
superimposed on a map of 2MASS and UCAC M-giants in the \sgr~ region
\citep[see][]{majewski03}.  

Concerning the pre-reduction, we applied to each frame overscan correction, bias subtraction and trimming
according to  the ESO-VIMOS
pipeline. 
Fringing correction on I band frames was performed using IRAF tasks {\tt
imcombine} and {\tt ccdproc}.


\section{Data analysis}

\subsection{VIMOS Photometry}

Photometry has been performed using the DAOPHOT/ALLFRAME packages
\citep{stetson94}. Templates for the PSF have been
selected in a semi--interactive mode, having selected a sample of bright
stars uniformly distributed across the entire frame. 
The DAOPHOT task ``{\tt psf}'' has been used to model a first PSF which was used to perform a first photometry. 
The subsequent step has been to select from the catalogue so obtained a new sample of stars to model a more accurate PSF and to perform a second photometry on each individual frame.
The photometric catalogs of the same field have been matched with
DAOMATCH/DAOMASTER in order to perform the final
photometry simultaneously in all the frames using ALLFRAME.  DAOMASTER finally 
allowed us to obtain the catalogue for each field.  
The photometric calibration was performed using zero points obtained by
the ESO Quality Control Team\footnote{The measurements and values of
  the ZP for the observing nights of this program can be retrieved at
  the site http://www.eso.org/observing/dfo/quality/}. Color terms
have been derived through a set of standard stars observed during the same nights.
The following correction was finally applied to the raw VIMOS photometry:

\begin{equation}
M = 2.5*\log(Flux) - a*Colour + cext*air +Zp + ap.
\end{equation}

where {\it a} is the color term, {\it air} is the airmass, {\it cext}
is the extinction coefficient, {\it Zp} is the zero point and {\it ap} is the aperture correction.

\subsection{Internal consistency}
\label{redsec}

\begin{figure}
   \resizebox{\hsize}{!}{\includegraphics{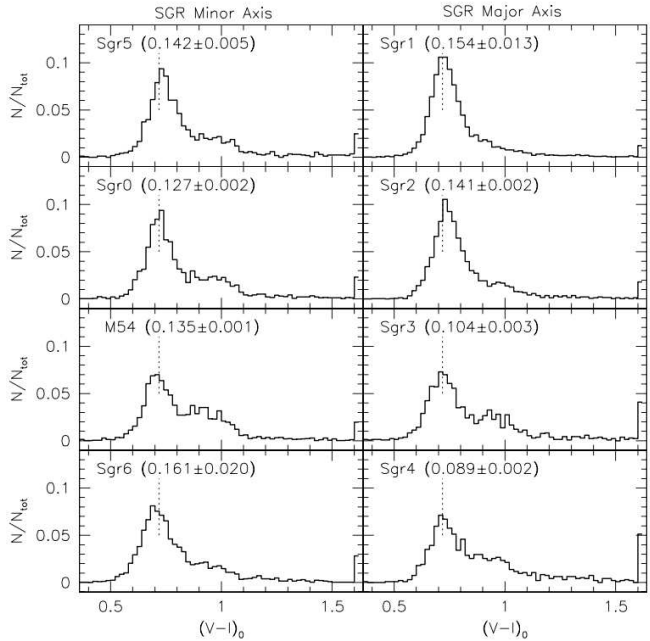}}
   \caption{Absolute (V$-$I)$_0$ color distribution of the MW
     sequence in the magnitude range $18.5<$V$<19.5$ for all the
     single fields. The magnitudes of each star has been de-reddened  using values from \citet[][]{schlegel98} corrected by \citet[][]{bonifacio00b}. The histograms are
     normalized to the total counts in the given magnitude range. The
     vertical dotted line shows the position of the maximum in the
     M~54 field, taken as a reference. In parenthesis the average value with rms scatter of
     E(B$-$V) for the given field.}
        \label{red}
   \end{figure}

In order to check the internal consistency of our calibration, we selected, in each of our
fields, a sample of bona-fide MW stars (stars in the interval $18.5<V<19.5$ and 
$0.8<V-I<1.05$) and corrected the catalogues according the maps of interstellar reddening of
\citet{schlegel98}\footnote{see http://www.astro.princeton.edu/~Schlegel/dust/}, as
corrected by \cite{bonifacio00b}. 
A comparison of the color distribution of these stars (see Fig. \ref{red}) reveals a good 
homogeneity of our data.


\subsection{VIMOS MOS spectroscopy}

In order to obtain spectra with good and stable calibrations
we implemented a new procedure to extract and calibrate VIMOS-MOS spectra and to derive 
reliable radial velocities, because the existing 
 ESO-VIMOS pipeline (ver. 2.0.15, see \citet[][]{cizzo}) produced
output spectra which turned out to be unsatisfactory both in terms of
extraction quality and wavelength calibration.  
Therefore we wrote an interactive procedure within the ESO
pipeline which allows fine tuning of several
parameters and better control of each calibration step.  
A typical VIMOS-MOS reduction 
consists of the following 5 steps:

\begin{enumerate}
\item  {\tt vmbias} to obtain a master bias frame;
\item  {\tt vmspflat} to obtain the master screen flat and 
       the not-normalised screen flat field;
\item  {\tt vmspcaldisp} to produce an extraction table;
\item  {\tt vmmosstandard} to obtain the spectro-photometric table;
\item  {\tt vmmosobsstare} (or {\tt vmmosobsjitter} in case of jittered science exposures) 
       to obtain the final reduced science exposure;
\end{enumerate}

While steps 1--2, 4--5 do not present particular difficulties, the key passage
is the 3rd step, i.e. the creation of a reliable extraction table.  In our case the spectral distortions models, described in
the input FITS headers, did not
guarantee an accurate identification of the reference arc lamp lines and
of the flat field spectra edges. Consequently we created a procedure to modify six recipe parameters until a good
extraction table was obtained.
According to our procedure the MOS\_SCIENCE\_SKY frame, containing the sky modeled
for each slit spectrum, is used to perform the reduction quality check and derive for five selected skylines 

\begin{equation}
   diff = (Sky_{obs} - Sky_{exp} )
\end{equation}

(where $Sky_{obs}$ is the observed wavelength of the sky line and
$Sky_{exp}$ is the expected wavelength) for each line and for each
slit.  
The accuracy in wavelength of the calibration is then estimated
from the residuals.
Typical values are of the order of 0.2 pixels.

Radial velocity of each star has been determined
using the three lines of the Ca triplet (849.8, 854.2 and 866.2 nm)
Typical uncertainty of
the radial velocities obtained in this way varies from 10 to 20 km/s.

\section{Comparison with previous photometry}

As no overlap with data in literature exists for \sgr~ peripheral fields,
we used the photometry of the field around \mfifo~ to check our
calibration.  
In particular we used the photometry of \citet{LS00}  (henceforth  LS00) who worked on an area of
11.1 $\times$ 11.1 arcmin centered on \mfifo.
In addition we used the standard stars of \citet{stetson00} around \object{NGC 4147} as his 
field partly overlaps with our observations in VIMOS quadrant 2.

\begin{figure}[t]
   \resizebox{0.95\hsize}{!}{\includegraphics{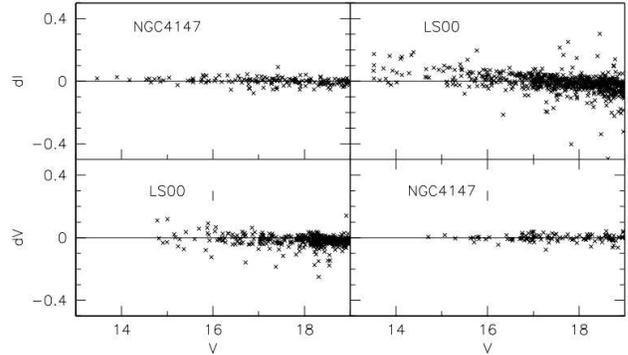}}
   \caption{Comparison between our
     photometry and references photometries. Difference in magnitude in
     V and I bands is plotted versus instrumental magnitudes.  LS00 is
     \cite{LS00} and \object{NGC 4147} are Stetson's standard stars.}
         \label{confvi}
   \end{figure}

\begin{figure}[ht]
   \resizebox{0.95\hsize}{!}{\includegraphics{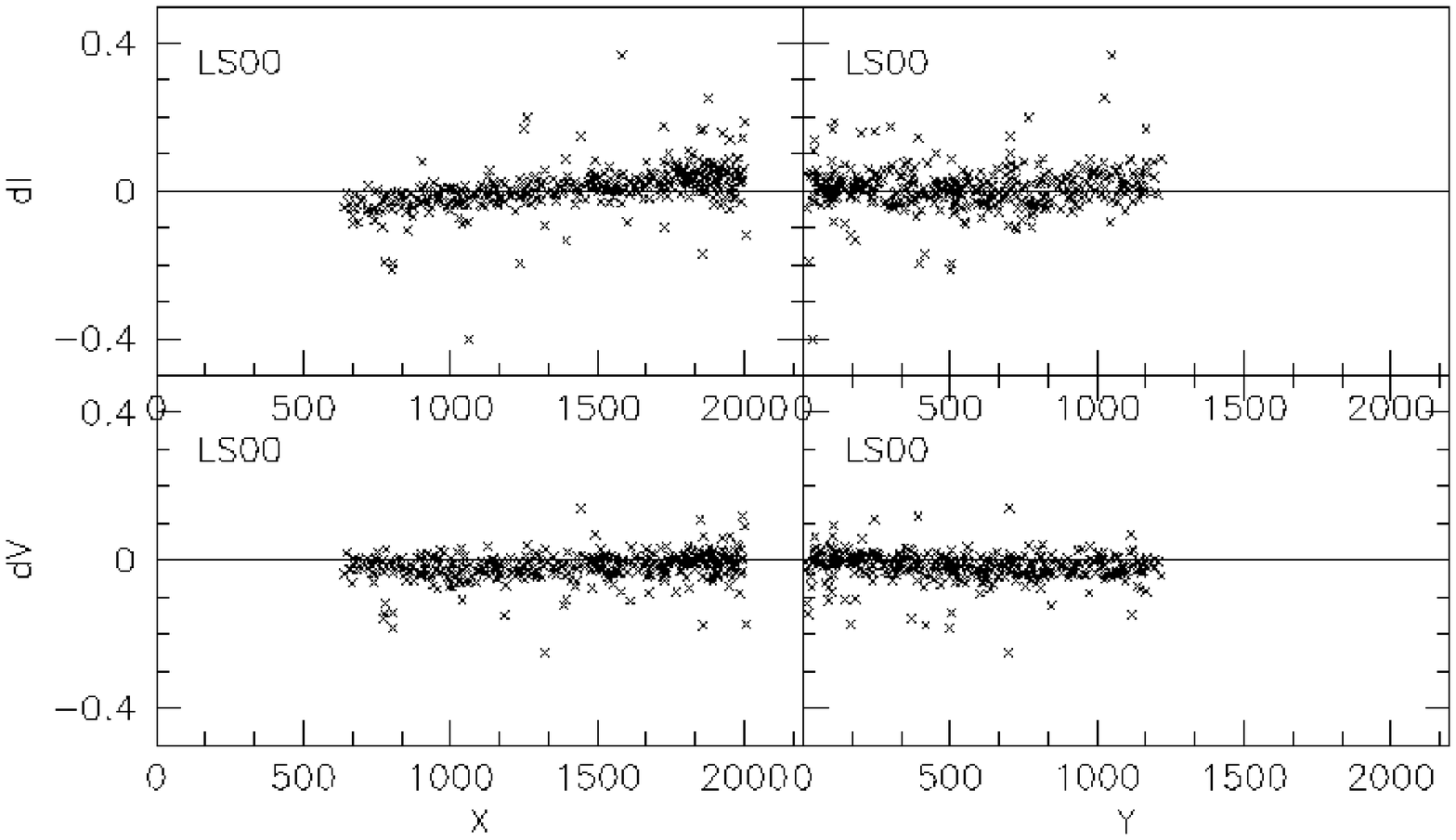}}
   \caption{Comparison between our
     photometry and reference photometry LS00. Difference in magnitude
     in V and I bands is plotted versus X and Y coordinates in pixel of
     the stars.  Only stars with V$<19$ or I$<19$ have been plotted
     LS00 is \cite{LS00}}
        \label{confxylay}
   \end{figure}

\begin{figure}[ht]
   \resizebox{0.95\hsize}{!}{\includegraphics{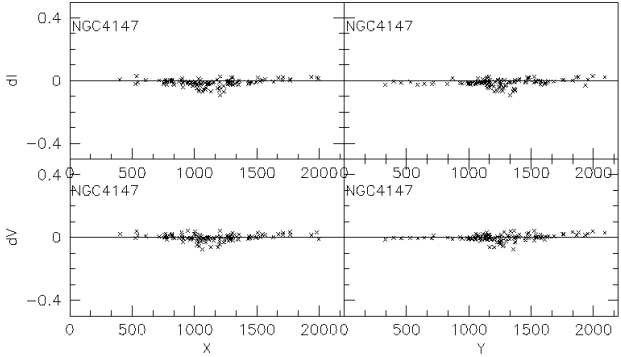}}
   \caption{Comparison between our
     photometry and reference photometry on \object{NGC 4147}. Difference in
     magnitude in V and I bands is plotted versus X and Y coordinates
     in pixel of the stars Only stars with V$<19$ or I$<19$ have been
     plotted. }
         \label{confxyste}
   \end{figure}

Figures \ref{confvi},  \ref{confxylay} and \ref{confxyste}
present the residuals between our photometry and the aforementioned studies for
stars with V$<19$ or I$<19$.  
While a significant dispersion is clearly visible for LS00, the comparison with NGC 4147 shows  
a good photometric agreement and the absence of obvious trends with the position of the 
stars (the root mean square for V data is about 0.025 with a sample of 109 stars, for I data is about 0.023 with a sample of 159 stars).


Actually two additional photometric studies exist, namely those of
\citet{M98} and \cite{M02}.
However considering that the first one covers
a small area of the sky (5.6 $\times$5.7 arcmin) and that the second one is based on WIFI@ESO2.2 data which 
is known to suffer of significant inhomogeneity \citep[][]{koch04,calamida08} we didn't consider useful to perform
such comparisons.
In conclusion, of the two studies we compared with, one (Stetson) supports our calibration while the other (LS00) presents
trends both in position and in magnitude which are impossible to disentangle at the moment. All in all 
the comparisons did not offer particular reasons to question our calibration.

\section{Color  Magnitude Diagrams}

\begin{figure*}
  \includegraphics[width=17cm]{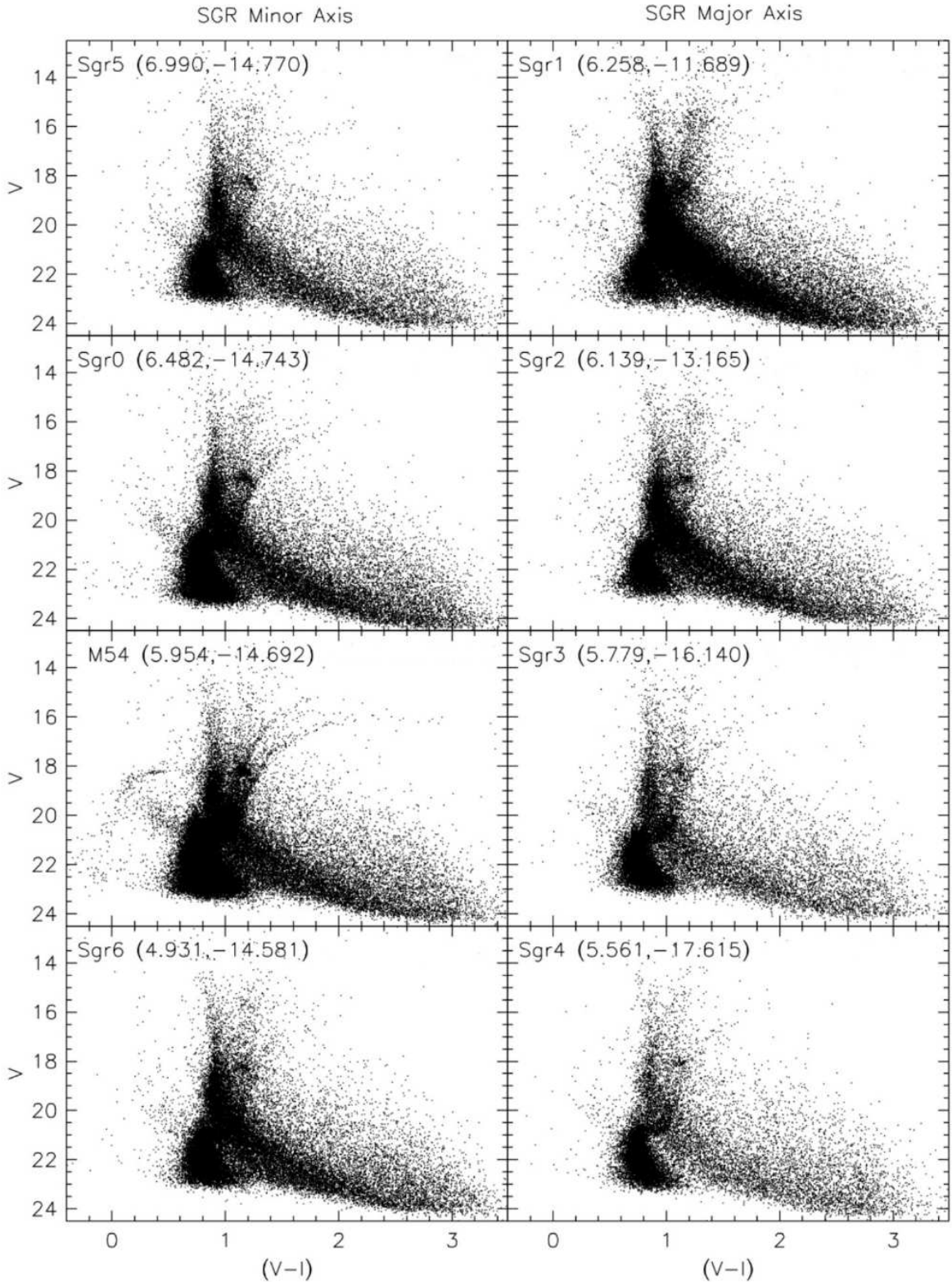}
  \caption{VIMOS Colour-Magnitude Diagrams (CMDs) for all observed field. 
    Each box shows the name of the field and the galactic coordinates in
    parenthesis. {\bf Left column:}
    CMD of the fields on the center (\object{M54}) and along the Minor Axis of
    SGR.  {\bf Right column:} CMDs of the fields along the Major Axis
    of SGR.}
  \label{cmdall}
\end{figure*}

In Fig. \ref{cmdall} we present the CMDs of the seven \sgr~ peripheral
fields plus the field centered on \mfifo~. Fields along the minor and major 
axis are presented in the left and in the right panel respectively.
Inspection of the CMDs reveals immediately that:

\begin{itemize}

\item contamination of the disk of the Milky Way is visible as a roughly  vertical
  sequence at V$-$I$\simeq0.8$ and from V=20.5 to V=13.5.  The giant branch of
  the old Bulge population is visible as a second sequence,
  nearly parallel to the previous one, at V$-$I$\simeq1.0$.  

\item  A very similar galactic 
 contamination can be observed in all the minor axis fields, which are 
 centered at nearly constant galactic latitude. The fields along the major axis
   show a galactic
  contamination which dramatically increases from field Sgr~4 to field Sgr~1;

\item the turn off of the Sgr main population is clearly detectable in
  all fields at a magnitude between V=20.5 and V=21.5 and color V$-$I$\simeq0.8$.

\item a densely populated red clump at V$\simeq18.0$ and
  V$-$I$\simeq1.2$ is detectable in all fields;

\item the Red Giant Branch (RGB) of \sgr~ main population is
  clearly present in field Sgr 2, 3, 5, 6 and in the \mfifo~ field. In
  Sgr 1 and Sgr 4 the number of RGB stars is lower, but the RGB
  remains nevertheless detectable;

\item an extended Blue Plume (BP) is visible in all fields in the intervals
  $0.4<$V$-$I$<0.8$ and $19.0<$V$<21.0$.  This population, which is
  clearly visible in the fields \mfifo, Sgr 0, 5, and 6, is considered to
  be the Main Sequence of the younger, metal rich population of
  \sgr~\citep[][]{bonifacio00,bonifacio2004,siegel07,sbordone2007}.
  The alternative interpretation of the plume as a sequence of Blue Stragglers 
  cannot be ruled out  \cite[see
  discussion in][]{Momany07}.

\end{itemize}

\section{Sgr dSph main population }

All the \sgr~ fields are characterized by the presence of a dominant
population whose progeny is the ubiquitous RGB .  To investigate the
possible presence of population gradients along the major and minor
axes we carried out a comparison of the position of the RGB in each field.

   \begin{figure}
   \resizebox{\hsize}{!}{\includegraphics{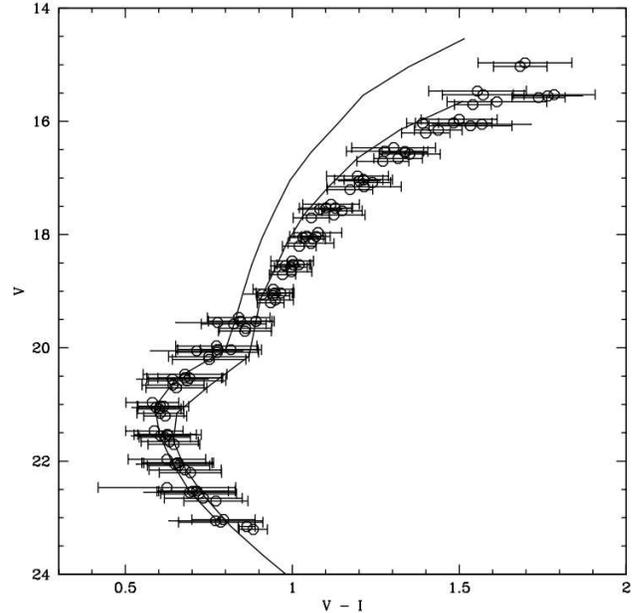}}
   \caption{\sgr~  main population position on \object{Sgr dSph} fields with uncertainties in color.  
   Fiducial lines of 2 galactic globular cluster are
     superimposed: \object{M5} and \object{47Tuc} characterized by
     [Fe/H]=-1.24, - 0.67}
        \label{sgrmainfid}
   \end{figure}


After having derived fiducial line of the main population in each field and having corrected for reddening differences, we
plotted all the fiducial lines in Fig. \ref{sgrmainfid} as open circles (Sgr1 was not reported because the high contamination 
of the MW did not allow to draw a reliable fiducial line).
RGB positions, used for the fiducial lines, and associated uncertainties are calculated finding the median value of the color distribution of
stars on each magnitude interval around the expected position of the
RGB, SGB and main sequence, and fitting with a gaussian the distribution
 around the median value.
On the same figure the fiducial lines of \object{M5} and \object{47Tuc} are drawn.
A simple inspection of Fig. \ref{sgrmainfid} reveals that:

\begin{enumerate}
\item Within the uncertainties the dominant population of \sgr~ is similar across the field.
\item Metallicity of \sgr~ main population is comparable with that of \object{47Tuc} 
([Fe/H]= -0.67), confirming previous estimates; 

\end{enumerate}

\section{Looking for other populations}
\label{fid}

\begin{figure}
\centering
   \resizebox{\hsize}{!}{\includegraphics{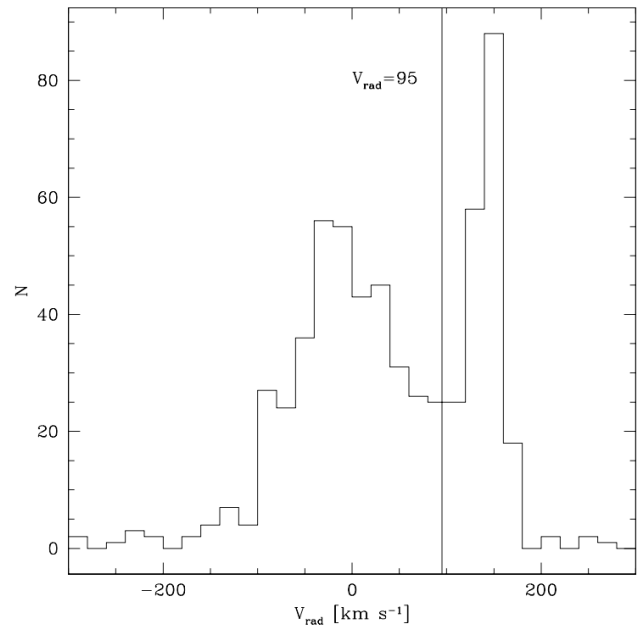}}
   \caption{Radial velocity distribution of the FLAMES sample. All stars with $V_{rad}>95$ km/s 
   are candidate \object{Sgr dSph} members}
     \label{allflames}
\end{figure}

\begin{figure}
\centering
   \resizebox{\hsize}{!}{\includegraphics{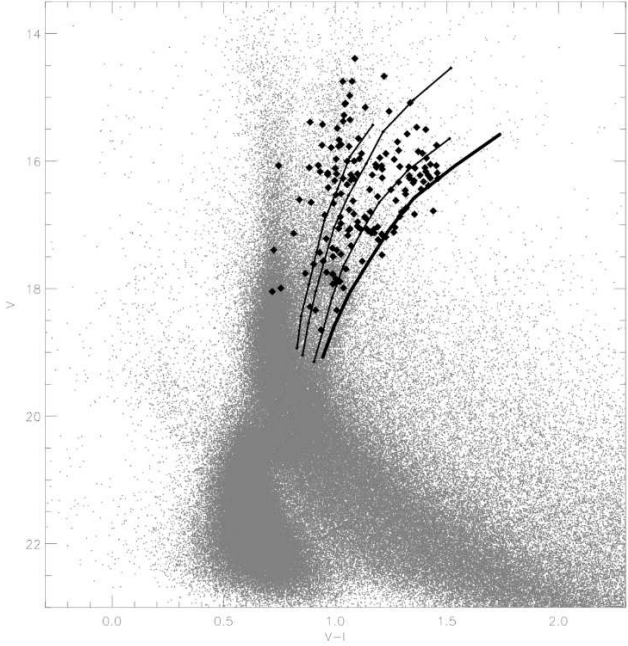}}
   \caption{Colour-magnitude diagram for all the stars detected on the
     7 \object{Sgr dSph} peripheral fields with overplotted the stars observed with
     FLAMES and showing radial velocities compatible with a membership
     in \sgr. Fiducial lines of 3 galactic globular cluster are
     superimposed.
     From left to right : \object{M92}, \object{M5} and \object{47Tuc} characterized by
     [Fe/H]=-2.52, -1.24, -0.67.
     The bold line follows the position of the RGB of \object{Sgr dSph} main population.}
     \label{cmdgala}
\end{figure}

From VIMOS-MOS observations we acquired spectra for 1000 stars, and we selected 180
 candidate \sgr ~ 
members from radial velocities centered on the 7 \sgr ~ fields.
575 targets have been added to this sample selecting other stars near 
in color and magnitude to the 180 confirmed Sgr dSph members.
Finally FLAMES followup has been performed on these stars.
After having measured radial velocities for these stars
we plotted these velocities in the histogram of Fig. \ref{allflames}. We have then selected the sample with $V_{rad} > 95$ 
because we considered that the large majority of this sample 
is constituted by \object{Sgr dSph} stars. The full analysis of the FLAMES sample  will be  presented in Sbordone et al., 2009.
The stars selected in this way are reported as black dots in Fig. \ref{cmdgala}, in addition to the complete photometry
of the present paper and with the 
the superimposed fiducial lines of \object{M 92}, \object{M 5} and \object{47 Tuc} (based on the photometry of
\citep{stetson00}).
Inspection of Fig. \ref{cmdgala} reveals that the sample selected on the basis of radial velocity span
a large interval of metallicity,  with stars more metal rich
than \object{47 Tuc} ([Fe/H]=-0.67), and stars more metal poor than \object{M92} ([Fe/H]=-2.52). 
These results
are in agreement with the data collected on the \object{Sgr dSph} core
\citep{zaggia2004, monaco2005, mcwilliam05, sbordone2007}, with two
remarkable exceptions, namely the presence of a well populated intermediate
population ([Fe/H] around -1) and a large numbers of bright 
stars lying at the blue edge of the RGB
(17$<V<14$ and 0.9$<V$-$I<1.1$) that cannot be reproduced with a
``\object{M 92} like'' population.  The case of these stars will be
analyzed in the next section.

\begin{table}
\caption{V - (V$-$I) fiducial lines for the globular 
    clusters \object{M 92}, \object{M 5}, \object{47 Tuc}, 
  . \label{fiduciMW}}
\centering
\begin{tabular}{lllll}
\hline
V    & V$-$I   & V$-$I   & V$-$I   & V$-$I  \\
mag  & \object{M 92} & \object{M 5} & \object{47 Tuc}  \\
\hline
12.0 &       & 1.557 & 1.566 &       \\
12.5 &       & 1.389 & 1.378 &       \\
13.0 & 1.180 & 1.253 & 1.251 &       \\
13.5 & 1.077 & 1.178 & 1.165 &       \\
14.0 & 1.023 & 1.098 & 1.090 &       \\
14.5 & 0.963 & 1.034 & 1.039 &       \\
15.0 & 0.928 & 0.992 & 1.000  \\
15.5 & 0.894 & 0.952 & 0.959 \\
16.0 & 0.859 & 0.919 & 0.942  \\
16.5 & 0.841 & 0.893 & 0.925 \\
17.0 & 0.815 & 0.869 & 0.812  \\
17.5 & 0.783 & 0.840 & 0.709  \\
18.0 & 0.638 & 0.694 & 0.699 \\
18.5 & 0.564 & 0.630 & 0.735 \\
19.0 & 0.571 & 0.644 & 0.795 \\
19.5 & 0.605 & 0.683 & 0.867 \\
20.0 & 0.653 & 0.734 & 0.964 \\
20.5 & 0.714 & 0.814 & 1.067 \\
21.0 & 0.788 &       &        \\
21.5 & 0.899 &       &        \\
22.0 & 0.957 &       &        \\
22.5 &       &       &        \\
\hline
\end{tabular}
\end{table}

\begin{figure}
\centering
   \resizebox{\hsize}{!}{\includegraphics{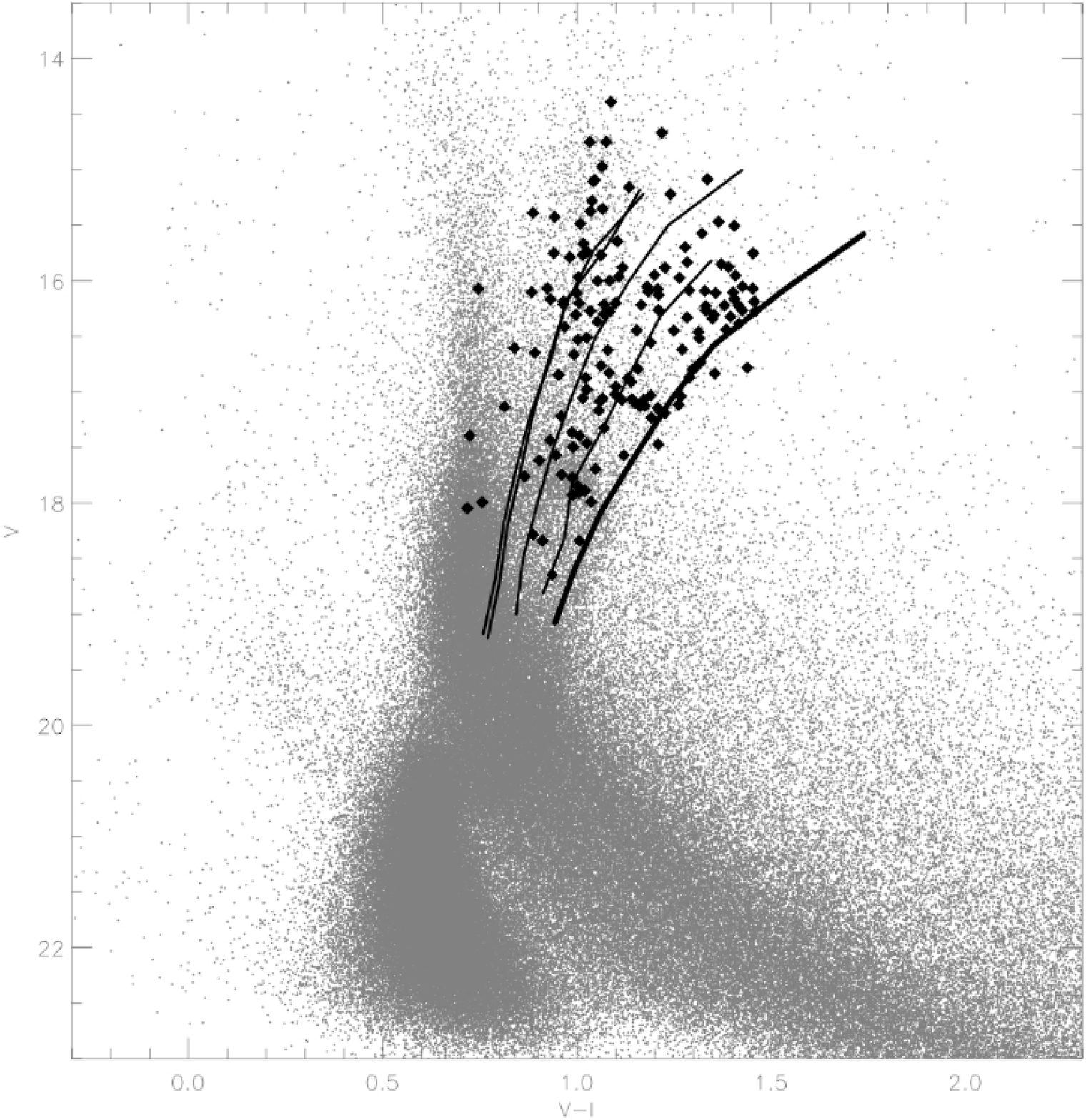}}
   \caption{Colour-magnitude diagram for all the stars detected on the
     7 \object{Sgr dSph} peripheral fields with overplotted the stars observed with
     FLAMES and showing radial velocities compatible with a membership
     in \sgr. Fiducial lines of 4 \sgr~ globular clusters are superimposed.
     From left to right : \object{Ter8}, \object{Arp2},
     \object{M54}, \object{Ter7} characterized by [Fe/H]=-2.34, -1.83, -1.55, -0.6.
      The bold line follows the position of the RGB of \object{Sgr dSph} main population.}
     \label{cmdsgr}
\end{figure}

The comparison of the fiducial lines for \sgr~
globular clusters \mfifo, \ters, \tere, and \arpt~  in Fig. \ref{cmdsgr} reveals a similar
scenario: \ters~ is not sufficiently metal rich to trace the most
metal rich \sgr~ population and the \tere~ fiducial line is not compatible with 
 the position of the bright stars at the blue edge of the RGB.

\begin{table}
\caption{V - (V$-$I) fiducial lines for the globular clusters 
\object{M 54}, \object{Terzan 7}, \object{Terzan 8}, \object{Arp 2}. 
\label{fiduciSgr}}
\centering
\begin{tabular}{llllll}
\hline
V    & V$-$I   & V$-$I   & V$-$I   & V$-$I  \\
mag  & \object{M 54} & \object{Ter 7} & \object{Ter 8} & \object{Arp 2} \\
\hline
15.5 & 1.631 &  ---  &  ---  &  ---  \\
16.0 & 1.440 & 1.425 & 1.349 & 1.299 \\
16.5 & 1.340 & 1.294 & 1.222 & 1.215 \\
17.0 & 1.254 & 1.219 & 1.150 & 1.109 \\
17.5 & 1.197 & 1.148 & 1.104 & 1.072 \\
18.0 & 1.145 & 1.067 & 1.064 & 1.019 \\
18.5 & 1.104 & 1.047 & 1.032 & 0.981 \\
19.0 & 1.067 & 0.992 & 0.999 & 0.948 \\
19.5 & 1.049 & 0.958 & 0.978 & 0.927 \\
20.0 & 1.010 & 0.933 & 0.948 & 0.896 \\
20.5 & 1.020 & 0.731 & 0.926 & 0.866 \\
21.0 & 0.806 & 0.660 & 0.755 & 0.781 \\
21.5 & 0.776 & 0.670 & 0.683 & 0.659 \\
22.0 & 0.794 & 0.701 & 0.670 & 0.672 \\
22.5 & 0.832 & 0.758 & 0.732 & 0.715 \\
23.0 & 0.879 & 0.816 & 0.774 & 0.778 \\
23.5 & 0.911 & 0.897 & 0.820 & 0.820 \\
\hline
\end{tabular}
\end{table}

\begin{table}
\caption{Reddening, Distance modulus and metallicity of the 3 Galactic globular clusters. 
\citep[][]{harris96, king98, carretta04, yong08} }
\centering
\begin{tabular}{lccc}
\hline
Object   & E(B$-$V)& $(m-M)_{V}$   & [Fe/H]  \\
\hline
M92      & 0.02  & 14.64 & -2.52 \\
M5       & 0.03  & 14.46 & -1.24 \\
47Tuc    & 0.04  & 13.37 & -0.67 \\
\hline
\end{tabular}
\end{table}

\begin{table}
  \caption{Reddening, Distance modulus and metallicity of the 4 \sgr~ globular clusters. \citep[][]{harris96, brown99, monaco04, sbordone2007, mottini08}} 
\centering
\begin{tabular}{lccc}
\hline
Object & E(B$-$V)& $(m-M)_{V}$ & [Fe/H]  \\
\hline
Ter8   & 0.12  & 17.45 & -2.34        \\
Arp2   & 0.10  & 17.59 & -1.83        \\
M54    & 0.15  & 17.10 & -1.55        \\
Ter7   & 0.07  & 17.05 & -0.60        \\
\hline
\end{tabular}
\end{table}

\subsection{RGB Blue edge}

\begin{figure}
\centering
   \resizebox{\hsize}{!}{\includegraphics{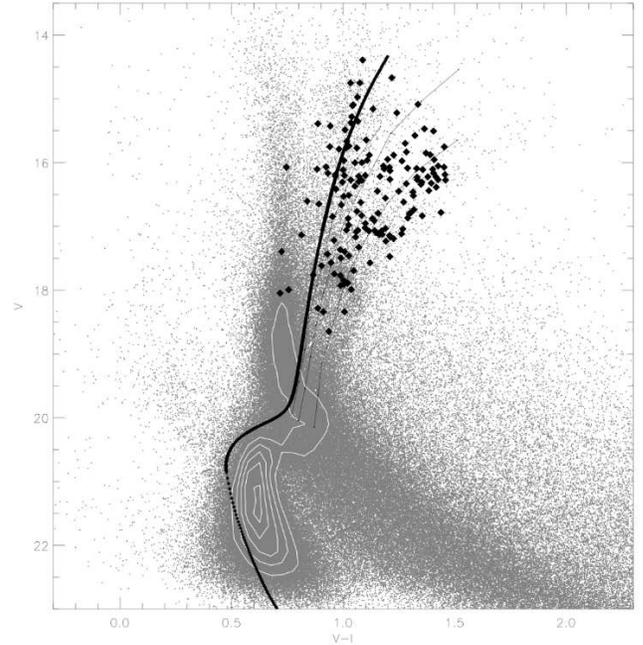}}
   \caption{Colour-magnitude diagram for all the stars detected on the
     7 \object{Sgr dSph} peripheral fields with overplotted the stars observed with
     FLAMES and showing radial velocities compatible with a membership
     in \sgr. Fiducial lines of the galactic globular clusters and
     isochrone characterized by 14 GYr [Fe/H]=-3.3 are superimposed. 
     Isodensity contours are superimposed on the \object{Sgr dSph} main sequence region.}
     \label{povera}
\end{figure}

In order to investigate on the nature of the 
stars in the region (17$<V<14$ and 0.9$<V-I<1.1$) 
we overplotted a 14 GYr [Fe/H]=-3.3 isochrone to the CMD (S. Cassisi, private
communication).
Inspection of Fig. \ref{povera}, however, rules out the 
hypothesis that the observed stars at the blue edge of the RGB
are representative 
of an extremely metal poor population, essentially for the anomalous expected position of
the main sequence and turn off.

\begin{figure}
\centering
   \resizebox{\hsize}{!}{\includegraphics{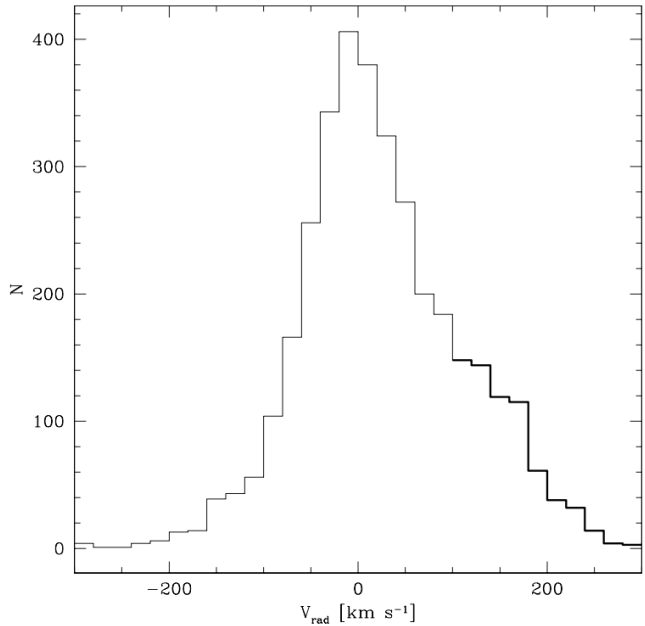}}
   \caption{Radial velocity distribution of the Besan\c con simulated
     sample in the ``RGB Blue edge'' box. The \sgr~ velocity
     compatible stars are plotted in bold.}
     \label{bulgevelbesa}
\end{figure}

On the other hand, in this region of the cmd a significant
contamination of the stars in the MW Bulge is expected, and the question is whether the
radial velocities we have measured may exclude they belong to the MW.
In order to clarify this point we have plotted in Fig. \ref{bulgevelbesa} the distribution
of radial velocities of the Besan{\c c}on sample \citep{robin04} for MW stars.
Inspection of Fig. \ref{bulgevelbesa} reveals immediately that the tail of the distribution, namely
about the 20 \% of the sample, fall in the 
region of the radial velocity of \sgr ~.
We can then conclude that the 
``RGB Blue edge stars'' are indeed heavily contaminated by Bulge stars and there is
the possibility that all of them are Bulge stars. It
cannot be excluded however that a fraction of them belongs to an extremely
metal poor \sgr~ population.

\section{Discussion and Conclusions}

In this paper we present a V, I VIMOS photometric catalog for
$\sim$320000 sources across eight fields in the direction of
the Sagittarius dwarf Spheroidal galaxy, covering roughly 0.43 square
degrees. The main purpose of this survey was to probe \sgr~ stellar
populations over a significant fraction of its spatial extension, from
its center to its periphery, and to provide targets for high
resolution spectroscopical analysis.

Our survey extends over 6.2 degrees along the galaxy major
axis, and 2 degrees along the minor axis, i.e. about one third of
\sgr~ extension. 
Two occurrences limit the extension of this sort of studies: first the presence
of the MW disk which makes  the contamination 
too severe at the position of our field Sgr 1, second the fact that  the density of \sgr~
stars drops quickly when moving away from the center, making a
spectroscopic follow-up highly inefficient. During the FLAMES
follow-up observations, in fields such as Sgr 1, Sgr4, Sgr 5, and Sgr
6 we were able to place only a few dozens of fibers on ``radial
velocity members'' detected by VIMOS-MOS. VIMOS spectroscopy provided
radial velocities adequate for FLAMES candidates selection, but not
good enough for dynamical analysis. The uncertainties in the MOS
wavelength calibration, together with the limited S/N of the spectra
led to a final uncertainty of about 10-20 km s$^{-1}$, comparable to the
overall \sgr~ velocity dispersion (see Fig. \ref{allflames}).

Within the precision allowed by the 
mentioned MW contamination, the population of  \sgr~
appears remarkably uniform among the observed fields. We thus confirm
the presence of a main \sgr~ population characterized roughly by the
same metallicity of 47 Tuc, but we also
found the presence of multiple populations on the peripheral fields
Sgr0-Sgr6, with a metallicity spanning from [Fe/H]=-2.3 to a nearly
solar value.  Our research  thus confirms
the extremely complex evolution
of this object.  Using our FLAMES data we plan to better
characterize these populations both from a chemical and a dynamical
point of view, adding new pieces to this intriguing puzzle.

\begin{acknowledgements}
The authors L.S. and P.B. acknowledge financial support from EU
contract MEXT-CT-2004-014265 (CIFIST). We thank S. Cassisi for
providing us an unpublished low metallicity isochrone.
This research has made use of the VizieR Service \citep{vizier} operated at CDS, Strasbourg, France, and of NASA's Astrophysics Data System.
\end{acknowledgements}


\end{document}